%% file: elbeprint2.tex
\documentclass[12pt]{article}
\usepackage{epsfig}

\newcommand\pubnumber{ANL-HEP-CP-01-002}
\newcommand\pubdate{\today}
\newcommand\hepnumber{hep-ph/0101164}

\def\support{\footnote{Work supported by the U.S. Department of 
Energy, High Energy Physics Division, under Contract No. W-31-109-Eng-38.}} 
\def\Title#1{\begin{center} {\Large\bf #1 } \end{center}}
\def\Author#1{\begin{center}{ \sc #1} \end{center}}
\def\Address#1{\begin{center}{ \it #1} \end{center}}

\newcommand\pubblock{\rightline{\begin{tabular}{l} \pubnumber\\
         \pubdate\\ \hepnumber \end{tabular}}}
\newenvironment{Abstract}{\begin{quotation}  }{\end{quotation}}
\newenvironment{Presented}{\begin{quotation} \begin{center} 
\end{center}
      \begin{center}\begin{large}}{\end{large}\end{center} \end{quotation}}
\def\Acknowledgments{\bigskip  \bigskip \begin{center}
          \large\bf Acknowledgments\end{center}}

\makeatletter
\def\section{\@startsection{section}{0}{\z@}{5.5ex plus .5ex minus
 1.5ex}{2.3ex plus .2ex}{\large\bf}}
\def\subsection{\@startsection{subsection}{1}{\z@}{3.5ex plus .5ex minus
 1.5ex}{1.3ex plus .2ex}{\normalsize\bf}}
\def\subsubsection{\@startsection{subsubsection}{2}{\z@}{-3.5ex plus
-1ex minus  -.2ex}{2.3ex plus .2ex}{\normalsize\sl}}

\renewcommand{\@makecaption}[2]{%
   \vskip 10pt
   \setbox\@tempboxa\hbox{\small #1: #2}
   \ifdim \wd\@tempboxa >\hsize     
       \small #1: #2\par          
     \else                        
       \hbox to\hsize{\hfil\box\@tempboxa\hfil}
   \fi}

 \def\citenum#1{{\def\@cite##1##2{##1}\cite{#1}}}
 
\newcount\@tempcntc
\def\@citex[#1]#2{\if@filesw\immediate\write\@auxout{\string\citation{#2}}\fi
  \@tempcnta\z@\@tempcntb\m@ne\def\@citea{}\@cite{\@for\@citeb:=#2\do
    {\@ifundefined
       {b@\@citeb}{\@citeo\@tempcntb\m@ne\@citea\def\@citea{,}{\bf ?}\@warning
       {Citation `\@citeb' on page \thepage \space undefined}}%
    {\setbox\z@\hbox{\global\@tempcntc0\csname b@\@citeb\endcsname\relax}%
     \ifnum\@tempcntc=\z@ \@citeo\@tempcntb\m@ne
       \@citea\def\@citea{,}\hbox{\csname b@\@citeb\endcsname}%
     \else
      \advance\@tempcntb\@ne
      \ifnum\@tempcntb=\@tempcntc
      \else\advance\@tempcntb\m@ne\@citeo
      \@tempcnta\@tempcntc\@tempcntb\@tempcntc\fi\fi}}\@citeo}{#1}}
\def\@citeo{\ifnum\@tempcnta>\@tempcntb\else\@citea\def\@citea{,}%
  \ifnum\@tempcnta=\@tempcntb\the\@tempcnta\else
  {\advance\@tempcnta\@ne\ifnum\@tempcnta=\@tempcntb \else\def\@citea{--}\fi
    \advance\@tempcnta\m@ne\the\@tempcnta\@citea\the\@tempcntb}\fi\fi}
\makeatother

\def\chargino{{\tilde{\chi}}^{\pm}}
\def\neutralino{{\tilde{\chi}}^{0}}
\def\gaugino{{\tilde{\chi}}}
\def\gluino{\tilde{g}}

\input econfmacros2.tex

\begin{document}
\begin{titlepage}
\pubblock

\vfill
\def\thefootnote{\fnsymbol{footnote}}
\Title{Next-to-leading Order Calculation \\[5pt] of Associated 
Production of Gauginos and Gluinos}
\vfill
\Author{Edmond L. Berger and T. M. P. Tait}
\Address{High Energy Physics Division\support, Argonne National Laboratory, 
Argonne, IL 60439,
USA\\E-mail: berger@anl.gov, tait@hep.anl.gov}

\Author{M. Klasen}

\Address{II.~Institut f\"ur Theoretische Physik, Universit\"at Hamburg, 
         D-22761 Hamburg, Germany\\E-mail: michael.klasen@desy.de} 
\vfill
\begin{Abstract}
Results are presented of a next-to-leading order calculation in
perturbative QCD of the production of charginos and neutralinos in
association with gluinos at hadron colliders.  Predictions for total 
and differential cross sections are shown at the energies of the Fermilab 
Tevatron and CERN Large Hadron Collider for a typical supergravity model of 
the sparticle mass spectrum and for a light gluino model.
\end{Abstract}
\vfill
\begin{Presented}
Presented by E. L. Berger at the \\ 
5th International Symposium on Radiative Corrections \\ 
(RADCOR--2000) \\[4pt]
Carmel CA, USA, 11--15 September, 2000
\end{Presented}
\vfill
\end{titlepage}
\def\thefootnote{\arabic{footnote}}
\setcounter{footnote}{0}

\section{Introduction}

The mass spectrum in typical supergravity and gauge-mediated models of 
supersymmetry (SUSY) breaking favors much lighter masses for gauginos 
than for squarks. Because the masses are smaller, there is greater 
phase space at the Tevatron and greater partonic luminosities for gaugino 
pair production, and for associated production of gauginos and gluinos, 
than for squark pair production.  Another point in favor of associated 
production is the relative simplicity of the final state.  For example, 
the lowest lying neutralino is the (stable) lightest supersymmetric 
particle (LSP) in supergravity (SUGRA) models, manifest only as missing 
energy in the events, and it is the second lightest in gauge-mediated models.  
The charginos and higher mass neutralinos may decay leptonically leaving a 
lepton signature plus missing transverse energy; relatively clean events ensue.  
Furthermore, associated production may be the best channel for measurement of 
the gluino mass.   

The search for direct experimental evidence of supersymmetry at colliders 
requires a good understanding of theoretical predictions of the total and 
differential cross sections for production of the superparticles.  In the case 
of hadron colliders, where collisions of strongly interacting hadrons are studied, 
the large strong coupling strength ($\alphas$) leads to potentially large 
contributions beyond the leading order (LO) in a perturbation series expansion 
of the cross section.  To have accurate theoretical estimates of production rates, 
it is necessary to include corrections at next-to-leading order (NLO) or beyond.  
In this contribution, we summarize our recent calculations at next-to-leading order 
in perturbative quantum chromodynamics (QCD) of the total and differential cross 
sections for associated production of gauginos and gluinos at hadron 
colliders\cite{letter,prdtext}.  Associated production offers a chance to study 
the parameters of the soft SUSY-breaking Lagrangian.  Rates are controlled by the 
magnitudes and phases of the gaugino ($\tilde \chi$) and gluino ($\tilde g$) 
masses and by mixing in the squark and gaugino sectors.  Our analysis is general 
in that it is not tied to a particular SUSY breaking model.  We can provide cross 
sections for arbitrary gluino and gaugino masses.
\begin{figure}[b!]
\vspace*{-2.0cm}
\begin{center}
\epsfig{file=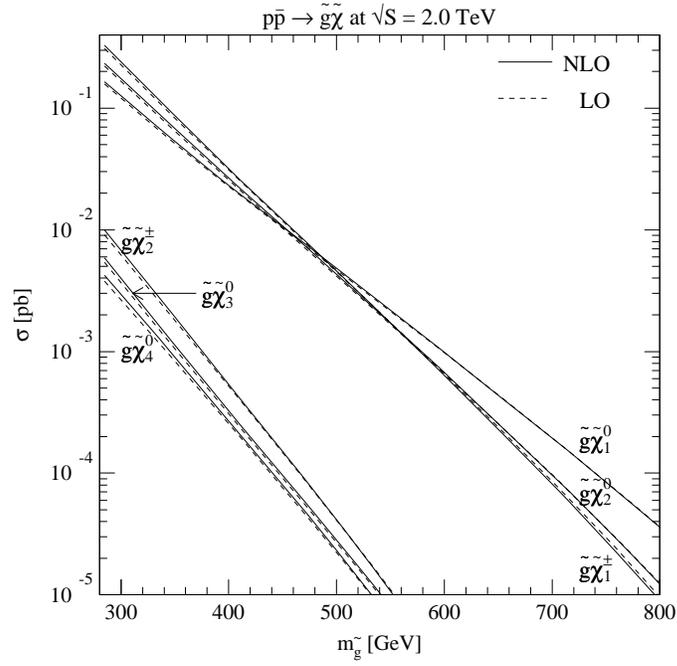,height=4in}
\epsfig{file=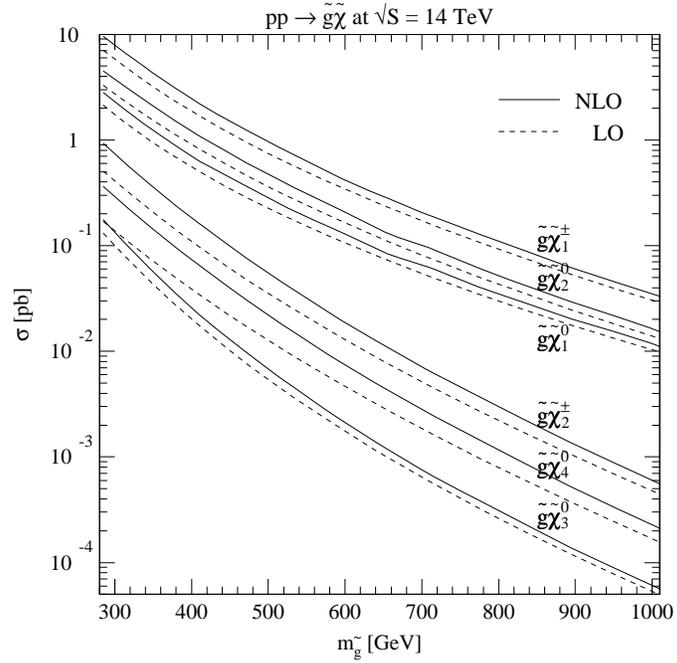,height=4in}
\caption{Predicted total hadronic cross sections at Run II of the Tevatron 
and at the LHC for all six $\tilde{g} \tilde{\chi}$ channels in a typical SUGRA 
model as functions of the gluino mass.}
\label{fig:xsecsugra}
\end{center}
\end{figure}

\section{NLO SUSY-QCD Formalism}
Associated production of a gluino and a gaugino proceeds in leading order
(LO) through a quark-antiquark initial state and the exchange of an intermediate 
squark in the $t$-channel or $u$-channel\cite{lo}.
At NLO, loop corrections must be included.  In addition, there are 
2 to 3 parton processes initiated either by quark-antiquark scattering, 
with a gluon radiated into the final state, 
$q + \bar{q} \rightarrow g + \gluino + \gaugino$, or by quark-gluon scattering
with a light quark radiated into the final state, 
$q + g \rightarrow q + \gluino + \gaugino$. For the quark-antiquark 
initial state, the loop diagrams involve the exchange of intermediate 
Standard Model or SUSY particles in self-energy, vertex, or box diagrams. 
Ultraviolet 
and infrared divergences appear at the upper and lower boundaries of 
integration over unobserved loop momenta.  They are regulated dimensionally 
and removed through renormalization or cancellation with corresponding 
divergences in the 2 to 3 parton (real emission) diagrams that have an 
additional gluon radiated into the final state.  In addition to soft 
divergences, real emission contributions have collinear divergences that 
are factored into the NLO parton densities.  

The set of Feynman diagrams with light quark emission includes 
diagrams in which an intermediate squark splits into a quark and either a 
gluino or gaugino.  After all initial state collinear singularities are removed 
by mass factorization, the matrix elements may still contain integrable singularities 
if the mass of the squark is larger than the mass of the gluino or gaugino.  In 
these cases, the intermediate squark state can be on its mass-shell.  These 
singularities represent the LO production of a squark and a gluino or gaugino, 
followed by the LO decay of the squark.  They may be regulated if one includes the 
full Breit-Wigner form for the squark propagator including a finite squark width.  
There is a further subtlety associated with the requirement that we not double-count 
the region of phase space in which the squark is on-shell.  The kinematic 
configuration with an on-shell squark is included in the LO production of
a squark and a gluino or a squark and a gaugino, and it should not be
considered as a genuine higher order correction to the production of gluinos
with gauginos.  To avoid double counting, we thus subtract the on-shell squark
contribution.  It can be subtracted leaving a genuine NLO contribution 
in the limit of small squark width.  
\begin{figure}[b!]
\begin{center}
\vspace*{-2.0cm}
\epsfig{file=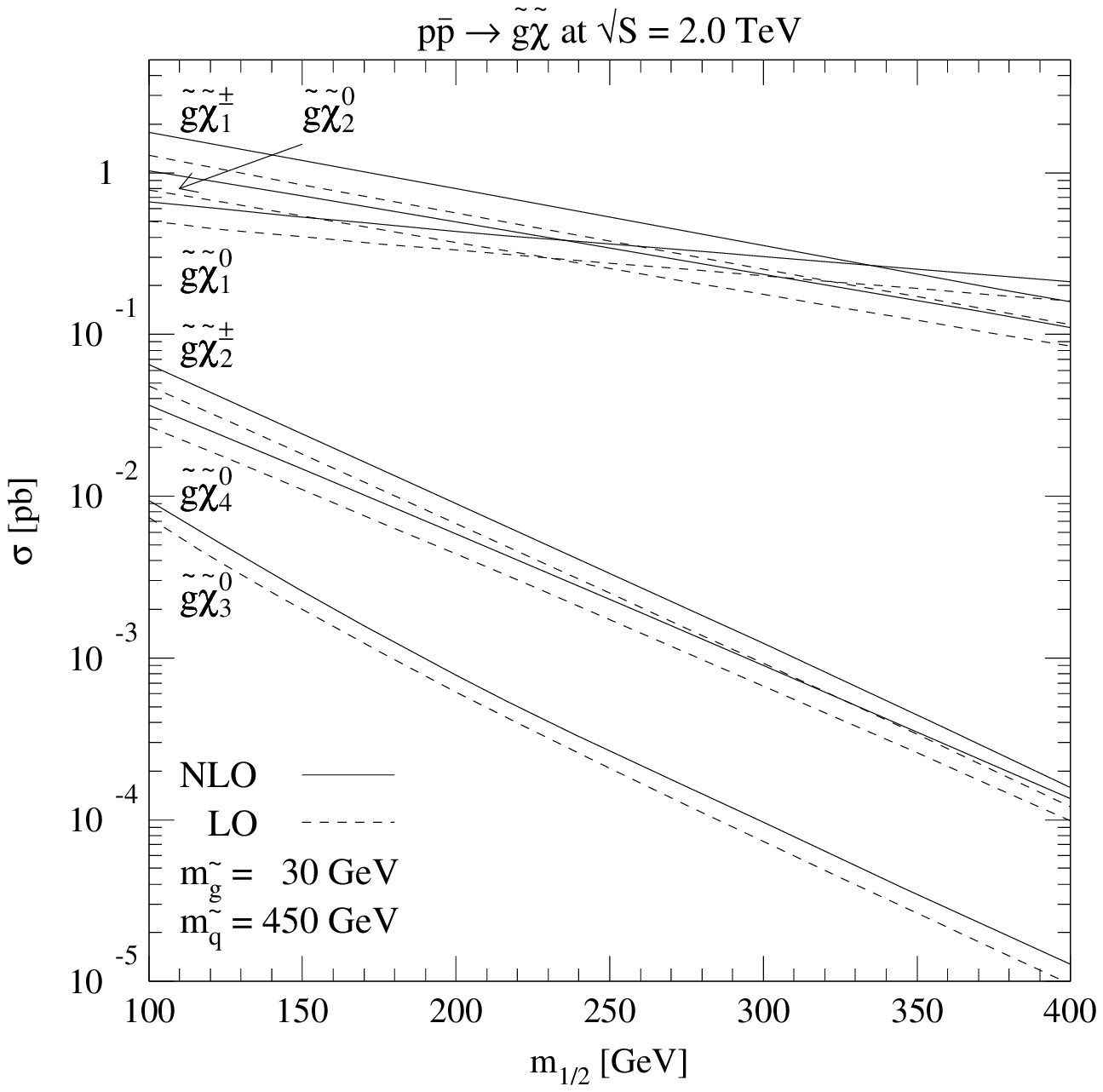,height=4in}
\epsfig{file=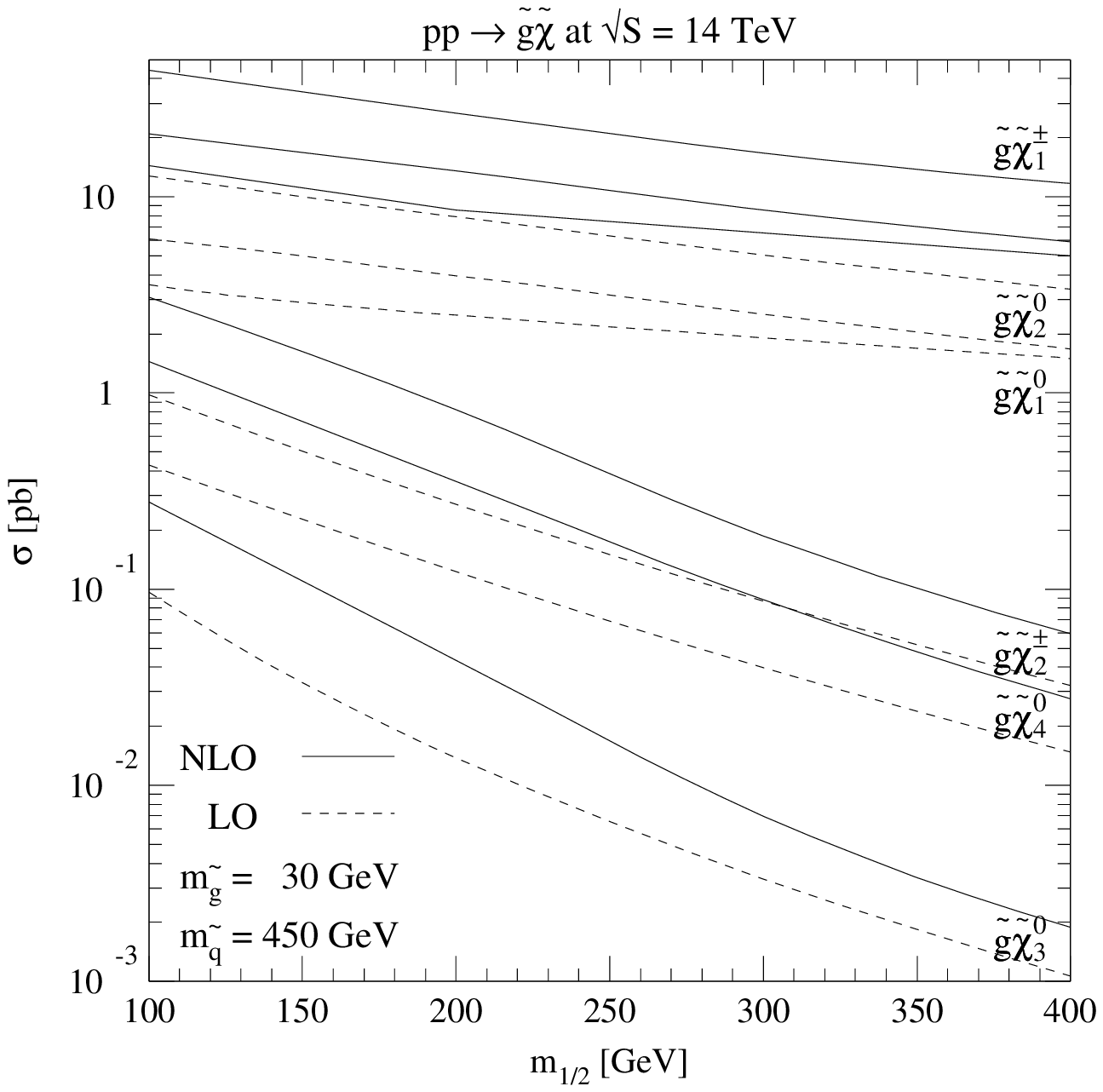,height=4in}
\caption{Predicted total hadronic cross sections at Run II of the Tevatron 
and at the LHC for all six $\tilde{g} \tilde{\chi}$ channels in our model 
with a gluino of mass 30GeV, as functions of the parameter $m_{1/2}$.}
\label{fig:xseclite}
\end{center}
\end{figure}

The full treatment of the NLO analysis is presented in our long paper\cite{prdtext}.

\section{Tevatron\,\,and\,\,LHC\,\,Cross\,\,Sections} 
To obtain numerical predictions for hadronic cross sections, we choose 
an illustrative SUGRA model with parameters $m_0=100$ GeV, $A_0$=300 GeV, 
tan $\beta$ = 4, and sign $\mu$ = +.  Because the gluino, gaugino, and 
squark masses all increase with parameter $m_{1/2}$ (but are insensitive 
to $m_0$), we vary $m_{1/2}$ between 100 and 400 GeV.  The resulting masses 
for $\tilde{\chi}_{1...4}^0$ vary between 31...162, 63...317, 211...665, 
and 241...679 GeV; $\tilde{\chi}_{1,2}^\pm$ are almost degenerate in 
mass with $\tilde{\chi}^0_{2,4}$. The mass $m_{\tilde{\chi}_3^0} < 0$ 
inside a polarization sum.  Our approach is general, and results can be 
obtained for any set of gaugino and gluino masses.  For our second model, 
we select one\cite{raby} with an intermediate-mass gluino as the 
lightest SUSY particle (LSP), fixing 
$m_{\tilde{g}} =$ 30 GeV, and $m_{\tilde{q}} =$ 450 GeV.  We choose 
a weak sector identical to the SUGRA case.  In our paper\cite{prdtext}, we also 
quote results for anomaly mediated, gauge mediated, and gaugino 
mediated models.

We convolve LO and NLO partonic cross sections with CTEQ5 parton densities 
in LO and NLO ($\overline{\rm MS}$)
along with 1- and 2-loop expressions for $\alpha_s$, the corresponding
values of $\Lambda$, and five active quark flavors.  

For the SUGRA case, we present total hadronic cross sections in 
Fig.\ \ref{fig:xsecsugra} as functions of the gluino mass.  The common 
renormalization and factorization scale $\mu$ is set equal to the average 
particle mass $m$.  The light 
gaugino channels should be observable at both colliders.  At the 
Tevatron, for $2~\rm{fb}^{-1}$ of integrated luminosity, 10 or more 
events could be produced in each of the lighter gaugino channels if 
$m_{\tilde g} < 450$ GeV.  The heavier Higgsino channels are suppressed 
by about one order of magnitude and might be observable only at the LHC.
As a rough estimate of uncertainty associated with the choice of 
parton densities, we note that the NLO cross section for 
$\tilde {\chi}_2^0$ production is lower by 12\% at the Tevatron 
with the CTEQ5 set than for the CTEQ4 set, and 4\% lower at the LHC.  
The impact of the NLO corrections can be seen more readily in the ratio 
of NLO to LO cross sections computed at a renormalizaton scale set equal 
to the average mass of the final state particles.  The NLO effects are 
moderate (of ${\cal O}$ (10\%)) at the Tevatron, while at the LHC the NLO 
contributions can increase the cross sections by as much as a factor of two. 
The second initial-state channel, initiated by gluon quark scattering, 
plays a significant role at the energy of the LHC. 

For the case of a gluino with mass 30 GeV, the total hadronic cross 
sections are shown in Fig.\ \ref{fig:xseclite} as functions of $m_{1/2}$.  
At the Tevatron, for $2~\rm{fb}^{-1}$ of integrated luminosity, 100 or 
more events could be produced in each of the lighter gaugino channels if 
$m_{1/2} < 400$ GeV.  In this case, NLO enhancement factors lie in the 
ranges 1.3 to 1.4 at the Tevatron and 2 to 4 at the LHC. 

\begin{figure}[b!]
\begin{center}
\vspace*{-2.0cm}
\epsfig{file=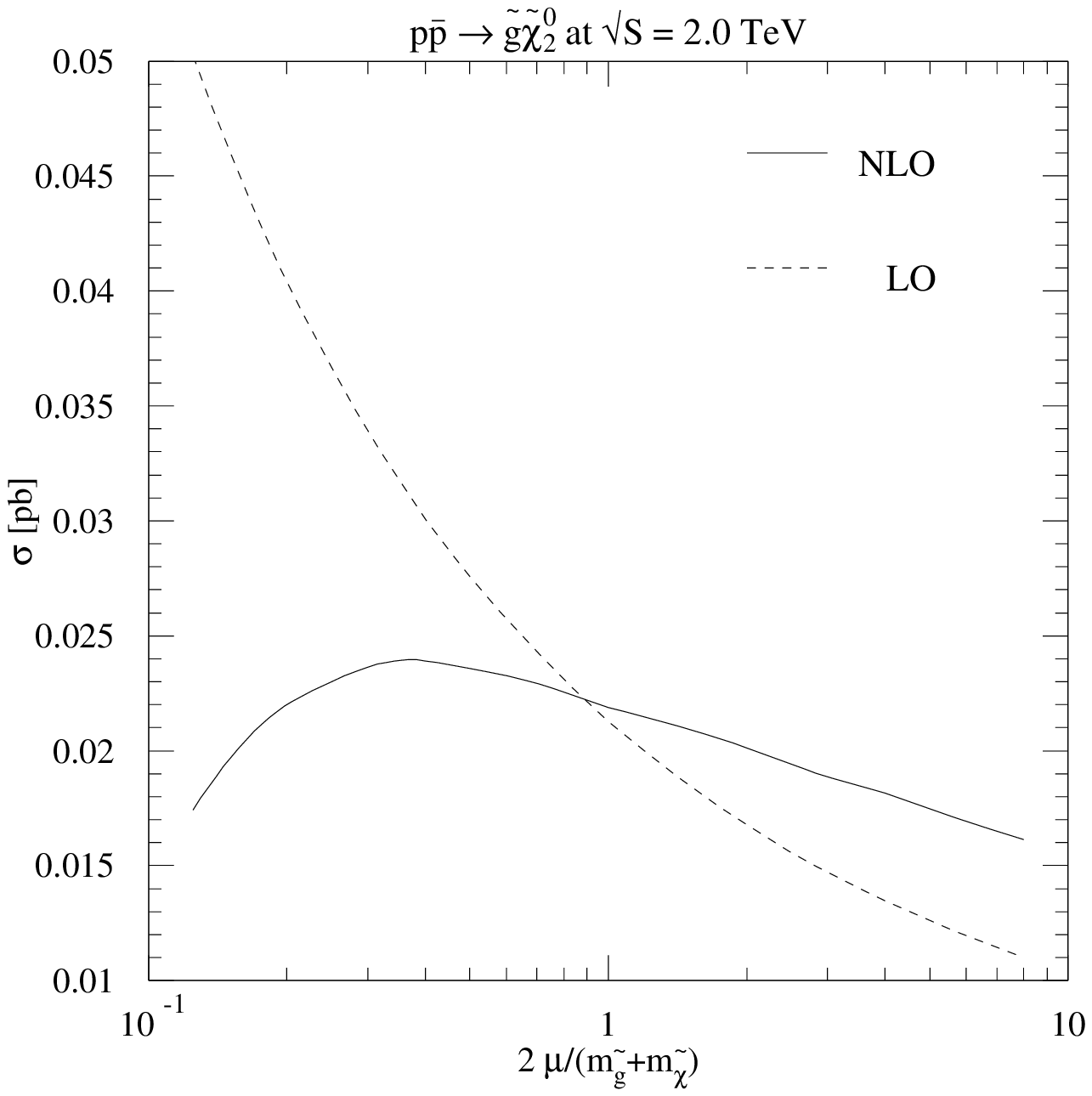,height=4in}
\epsfig{file=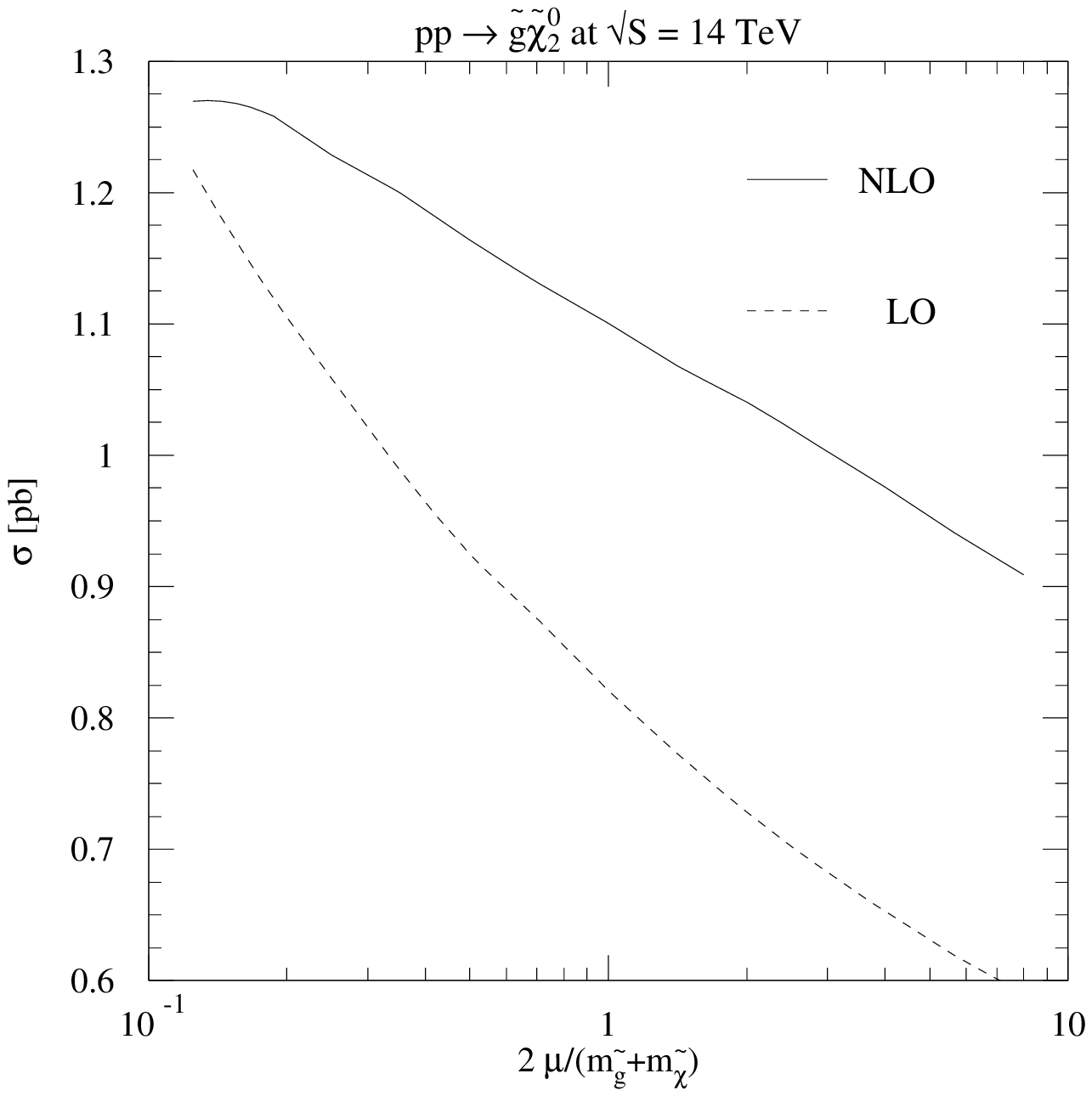,height=4in}
\caption{Dependence of the predicted NLO and LO total cross sections at the 
  Tevatron and at the LHC on the renormalization and factorization scale.  We show 
  the case of $\gluino \neutralino_2$ production in the SUGRA model, 
  with $m_{\gluino} = 410 $ GeV and $m_{\neutralino_2} = 104 $ GeV.}
\label{fig:mudep}
\end{center}
\end{figure}

An important measure of the theoretical reliability is the variation of 
the hadronic cross section with the renormalization and factorization 
scales.  At LO, these scales enter only in the strong coupling constant
and the parton densities, while at NLO they appear also explicitly in the 
hard cross section. The scale dependence is reduced considerably after NLO 
effects are included, as shown in Fig.\ \ref{fig:mudep}. The Tevatron (LHC) 
cross sections vary by $\pm 23 (12) \%$ at LO, but only by $\pm 8 (4.5) \%$ 
in NLO when the scale is varied by a factor of two around the central scale.

\begin{figure}[b!]
\begin{center}
\vspace*{-2.0cm}
\epsfig{file=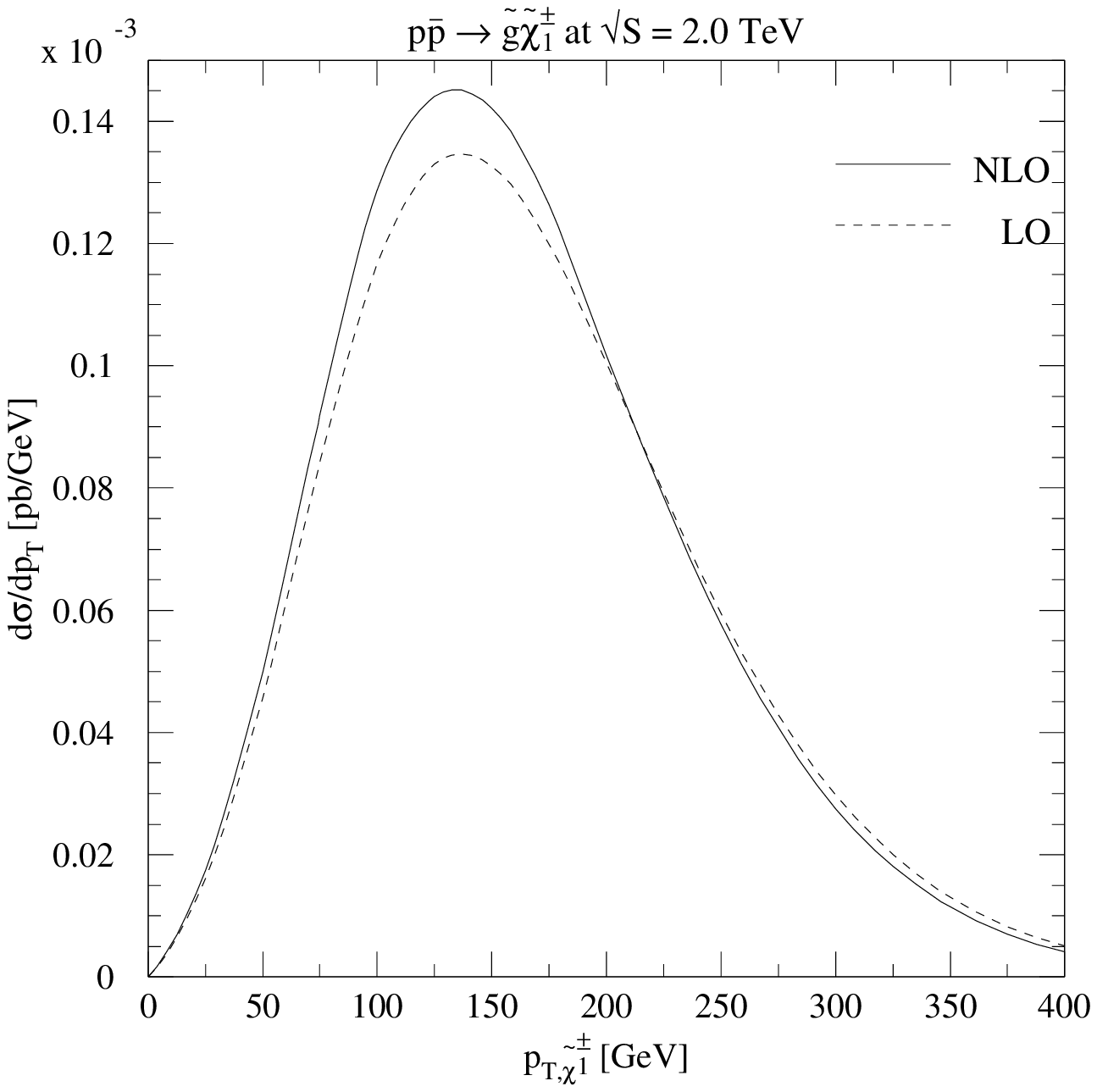,height=4in}
\epsfig{file=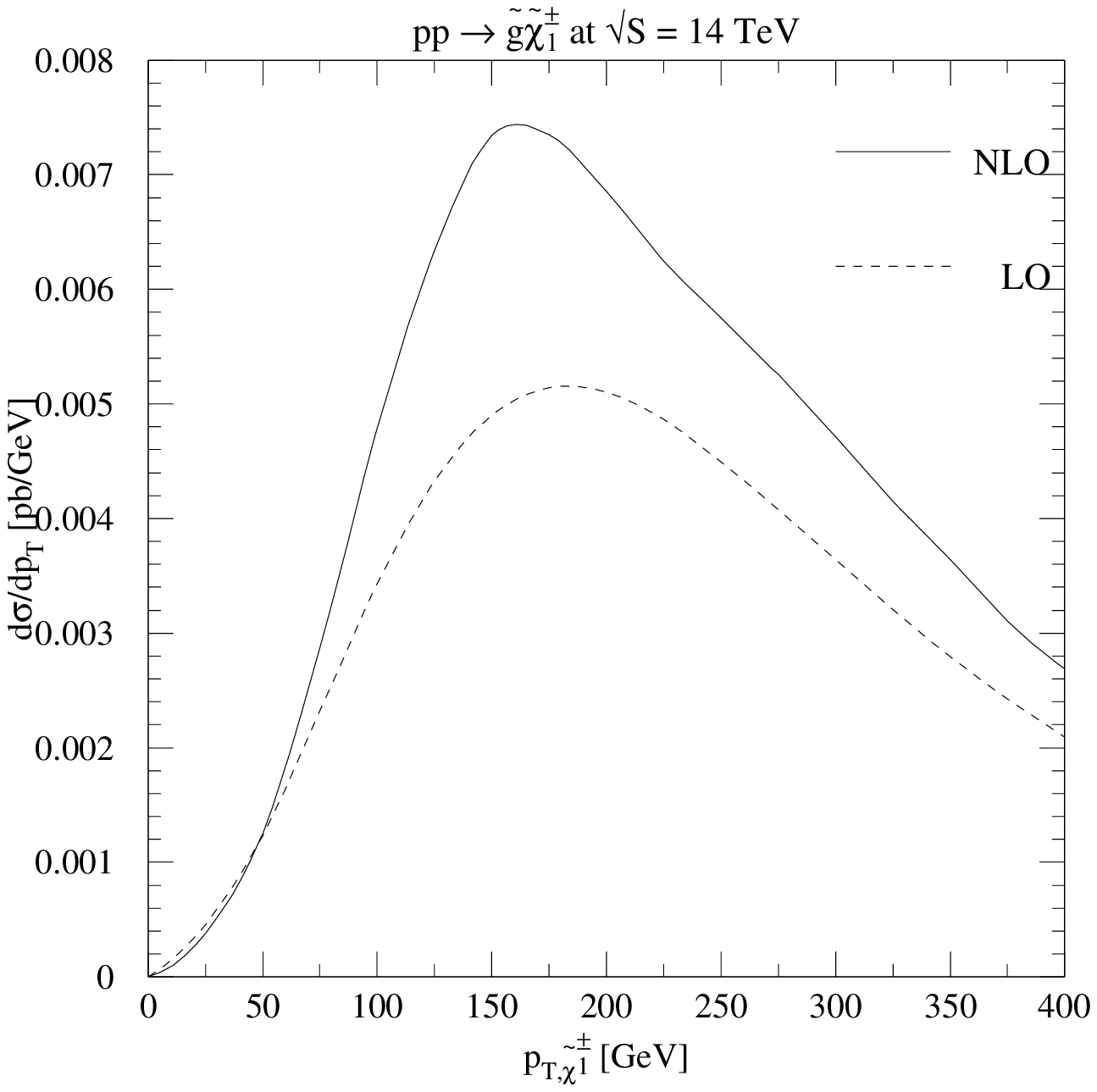,height=4in}
\caption{Differential cross section in transverse momentum $d\sigma/dp_T$ 
  for the production of $\chargino_1$ with mass 101 GeV in association with 
  a $\gluino$ of mass 410 GeV at the Tevatron and at the LHC.}
\label{fig:ptdep}
\end{center}
\end{figure}

For experimental searches, distributions in transverse momentum are 
important since cuts on $p_T$ help to enhance the signal.  In our 
long paper\cite{prdtext}, we show that NLO contributions can have a large
impact on $p_T$ spectra at the LHC, owing to contributions from
the $gq$ initial state.  At the Tevatron the NLO $p_T$-distribution is 
shifted moderately to lower $p_T$ with respect to the LO expectation.  
Examples for the $\tilde g \chargino_1$ channel are shown in 
Fig.\ \ref{fig:ptdep}.  The shapes of the rapidity distributions of the gauginos 
are not altered appreciably by NLO contributions.  

\section{Summary}

In our long paper we report a complete next-to-leading order analysis of
the associated production of gauginos and gluinos at hadron colliders.
If supersymmetry exists at the electroweak scale, the cross section for
this process is expected to be observable at the Fermilab Tevatron
and/or the CERN LHC. It is enhanced by the large color charge of the gluino
and the relatively small mass of the light gauginos in many SUSY models.
Associated production offers a chance to study in detail the
parameters of the soft SUSY-breaking Lagrangian.  The rates are proportional 
to the phases of the gaugino and gluino masses, and to the mixings in the 
squark and chargino/neutralino sectors.  In combination with other 
channels, associated production could allow one to measure some or all of 
these quantities.

The physical gluino and gaugino masses that we use, as well as the gaugino 
mixing matrices, are based on four popular SUSY breaking models plus a fifth  
scenario in which the gluino mass is relatively light.
Because the LO cross sections in gauge-mediated, gaugino-mediated, and 
anomaly-mediated supersymmetry breaking models are not too 
dissimilar from those of the SUGRA case at Tevatron energies, we focus our 
NLO work on the SUGRA model and on a model in which a light gluino, with mass 
$m_{\tilde g} =$ 30 GeV, is the lightest supersymmetric particle (LSP).  

In the SUGRA model, the largest cross sections at the 
Fermilab Tevatron energy are those for neutralino $\neutralino_2$, enhanced by 
its $\tilde{W}_3$-like coupling with respect to the $\tilde B$-like 
$\neutralino_1$, and the chargino $\chargino_1$, about 
equal in mass with the $\neutralino_2$. The NLO corrections to associated production 
are generally positive, but they can be modest in size, ranging in the SUGRA 
model from a few percent at the energy of the Tevatron to 100\% at 
the energy of the LHC, depending on the sparticle masses.  In the light-gluino 
case, NLO contributions increase the cross section by factors of 1.3 to 1.4 at 
the energy of the Tevatron and by factors of 2 to 3.5 at the energy of the 
LHC.  The large enhancements owe their origins to the important role of the $gq$ 
channel that enters first at NLO.  

Owing to the NLO enhancements, collider searches for signatures of associated 
production will generally discover or exclude sparticles with 
masses larger than one would estimate based on LO production rates alone.  
More significant from the viewpoint of reliability, the renormalization and 
factorization scale dependence of the cross sections is reduced by a 
factor of more than two when NLO contributions are included. 

At Run II of the Fermilab Tevatron, for an integrated luminosity of 
2 $\rm{fb}^{-1}$, we expect that 10 or more events could be produced in 
each of the lighter gaugino channels of the SUGRA model, 
$\gluino \neutralino_1$, 
$\gluino \neutralino_2$, and $\gluino \chargino_1$, provided that the 
gluino mass $m_{\gluino}$ is less than 450 GeV.  The cross sections 
for the three heavier gaugino channels, $\gluino \neutralino_3$, 
$\gluino \neutralino_4$, and $\gluino \chargino_2$, are smaller by 
an order of magnitude or more than those of the lighter gaugino 
channels.  In the light gluino LSP model, more than 100
events could be produced in the three lighter gaugino channels provided 
that the common GUT-scale fermion mass $m_{1/2}$ is less than 400 GeV, 
and as many as 10 events in the three heavier gaugino channels as long 
as $m_{1/2}$ is less than 200 GeV.  At the higher energy and luminosity 
of the LHC, at least a few events should be produced in every channel in 
the SUGRA model and many more in the light gluino model.    

The relatively large cross sections suggest that associated production is a 
good channel for discovery of a light gluino at the Tevatron, for closing the 
window on this possibility, and/or for setting limits on light gaugino masses.  
The usual searches for a light gluino LSP are based on the assumption 
that gluinos are produced in pairs.  In this situation, the dominant background 
is QCD production of hadronic jets.  Hard cuts 
on transverse momentum must be made to reduce this background to tolerable 
levels.  The cuts, in turn, mitigate against gluinos of modest mass.  By 
contrast, if light gluinos are produced in association with gauginos, one 
can search for light gluino monojets accompanied by leptons and/or missing 
transverse energy from gaugino decays.   

\Acknowledgments
Work is the High Energy Physics Division at Argonne National Laboratory is 
supported by the U.S. Department of Energy, High Energy Physics Division, under 
Contract No. W-31-109-Eng-38.
The work of M. Klasen is supported by Deutsche Forschungsgemeinschaft (DFG) under 
contract KL 1266/1-1.

\end{document}

%% file: econfmacros2.tex
\def\beq{\begin{equation}}
\def\eeq#1{\label{#1}\end{equation}}
\def\eeqn{\end{equation}}

\newenvironment{Eqnarray}%
   {\arraycolsep 0.14em\begin{eqnarray}}{\end{eqnarray}}
\def\beqa{\begin{Eqnarray}}
\def\eeqa#1{\label{#1}\end{Eqnarray}}
\def\eeqan{\end{Eqnarray}}

\let\bar=\overbar

\def\Dslash{\not{\hbox{\kern-4pt $D$}}}
\def\dslash{\not{\hbox{\kern-2pt $\del$}}}

\def\alphas{\alpha_s}
\def\msb{{\bar{\ssstyle M \kern -1pt S}}}

\def\lsim{\mathrel{\raise.3ex\hbox{$<$\kern-.75em\lower1ex\hbox{$\sim$}}}}
\def\gsim{\mathrel{\raise.3ex\hbox{$>$\kern-.75em\lower1ex\hbox{$\sim$}}}}